\documentclass[fleqn]{article}
\usepackage{amsmath,amssymb}
\usepackage{epsfig,graphicx}

\catcode`@=11 \@addtoreset{equation}{section} \catcode`@=12

\begin{document}
\title{\vskip-1.7cm \bf  BRST technique for the cosmological density matrix}
\date{}
\author{A.O.Barvinsky}
\maketitle
\hspace{-8mm} {\,\,\em Theory Department, Lebedev
Physics Institute, Leninsky Prospect 53, Moscow 119991, Russia}

\begin{abstract}
The microcanonical density matrix in closed cosmology has a natural definition as a projector on the space of solutions of Wheeler-DeWitt equations, which is motivated by the absence of global non-vanishing charges and energy in spatially closed gravitational systems. Using the BRST/BFV formalism in relativistic phase space of gauge and ghost variables we derive the path integral representation for this projector and the relevant statistical sum. This derivation circumvents the difficulties associated with the open algebra of noncommutative quantum Dirac constraints and the construction/regularization of the physical inner product in the subspace of BRS singlets. This inner product is achieved via the Batalin-Marnelius gauge fixing in the space of BRS-invariant states, which in its turn is shown to be a result of truncation of the BRST/BFV formalism to the ``matter" sector of relativistic phase space.
\end{abstract}

\maketitle
\section{Introduction}
The purpose of this paper is to justify the path integral representation for the statistical sum in quantum cosmology given by the integral over periodic metric and matter field configurations
    \begin{eqnarray}
    &&Z=
    \!\!\int\limits_{\,\,\rm periodic}
    \!\!\!\! D[\,g_{\mu\nu},\phi\,]\;
    e^{iS[\,g_{\mu\nu},\phi\,]}.         \label{Z}
    \end{eqnarray}
With the gauge fixing procedure \cite{DW,FP} implicit in the integration measure $D[\,g_{\mu\nu},\phi\,]$ this representation looks obvious. Very often, especially in context of quantum gravity on the lattice, it is simply postulated and serves as a starting point in the analysis of the UV limit, gravitational thermodynamics, etc. However, the principal point behind this object is a physical setting for the system with the action $S[\,g_{\mu\nu},\phi\,]$, that is a concrete definition of the quantum state corresponding to this statistical sum. In contrast to Euclidean quantum gravity studies on the lattice, where such questions are usually not posed, in applications to early quantum Universe this setting is critically important for the interpretation of (\ref{Z}).

Recently it was suggested that the path integral (\ref{Z}) represents the statistical sum of the microcanonical density matrix of spatially closed Universe given by the projector onto the space of solutions of Wheeler-DeWitt equations -- quantum Dirac constraints in gravity theory \cite{why}. If their collection -- operator realization of classical Hamiltonian and momentum constraints -- is denoted by $\hat H_\mu$ (condensed index $\mu$ including both their discrete labels and spatial coordinates), then formally the microcanonical density matrix generating  (\ref{Z}) was advocated in \cite{why} to be
    \begin{eqnarray}
    \hat\rho=\frac1Z\,
    \prod_\mu \delta(\hat H_\mu).       \label{projector}
    \end{eqnarray}

Conceptually this definition in {\em closed} cosmology was based on the fact that this is a natural generalization of the microcanonical density matrix in the usual non-gauge (and parametrization non-invariant) systems with a nonvanishing Hamiltonian $\hat H$, $\hat\rho\sim\delta(\hat H-E)$. This is because closed cosmology, in contrast to such systems, does not have nonvanishing global constants of motion -- global charges
like energy $E$, electric charge, etc. The basic set of constants of motion here is formed by local constraints on phase space of the theory $H_\mu=H_\mu(q,p)$, all having zero value. Application of this definition has recently led to the Euclidean quantum gravity path integral representation for this density matrix \cite{slih,bigboost} and to a number of interesting predictions. In particular, for the cosmologcal model driven by conformal field theory this has resulted in a limited range of the primordial cosmological constant, inflationary and dark energy scenarios \cite{bigboost}, thermal rather than vacuum nature of the generated CMB spectrum \cite{CMB}, etc.

All these conclusions were based on the calculation of the path integral (\ref{Z}). In its turn it was advocated to follow from the coordinate representation of the projector (\ref{projector}) in the form of the canonical path integral
    \begin{eqnarray}
    \langle\,q_+|\,\prod\limits_\mu\delta(\hat H_\mu)\,|\,q_-\rangle=
    \!\!\!\!\!
    \int\limits_{\,\,\,\,\,\,
    q(t_\pm)=\,q_\pm}
    \!\!\!\!\!\!\!\!\!
    D[\,q,p,N\,]\;
    \exp\left[\,i\!
    \int_{\,\,t_-}^{t_+} dt\,
    (p\,\dot q-N^\mu H_\mu)\,\right].   \label{rhocanonical}
    \end{eqnarray}
with the canonical ADM form of the gravitational action and the Faddeev-Popov gauge fixing procedure encoded in the integration measure $D[\,q,p,N\,]$. This is the integration over histories of gravitational phase space variables $(q,p)$ and Lagrange multipliers $N$ -- 3-metric coefficients and matter fields $q=(g_{ab}({\bf x}),\phi({\bf x}))$, their conjugated momenta $p$ and ADM lapse and shift functions $N^\mu=(N^\perp({\bf x}), N^a({\bf x}))$. These histories interpolate between the arguments $q_\pm$ of the projector kernel. The goal of this paper is to carefully derive this representation and the statistical sum (\ref{Z}) as the trace of (\ref{rhocanonical}), $Z={\rm tr}\,\prod_\mu\delta(\hat H_\mu)$.

Though the expression (\ref{Z}) seems natural, there is an obvious difficulty with its derivation from the definition (\ref{projector}). First, this definition is not formally consistent, because the operators $\hat H_\mu$ do not commute with each other and only form an open algebra. Second, taking the trace of the projector (\ref{projector}) should be done with respect to the physical inner product in the space of solutions of quantum Dirac constraints, which differs from simple integration over $q$ of the diagonal element of the kernel (\ref{rhocanonical}). In what follows we fill up these two omissions in the formalism of the theory. This will be done within the framework of the general BRST/BFV technique for quantum gauge systems, which allows one to specify precisely the integration measure $D[\,q,p,N\,]$, boundary conditions at $t_\pm$ on all integration variables and independence of the kernel (\ref{rhocanonical}) of the choice of $t_\pm$ (in accordance with the fact that the left hand side of (\ref{rhocanonical}) is $t_\pm$-independent, and the ``time" on the right hand side playing the role of operator ordering parameter).

In Sect.2 we begin with a brief overview of the BFV formalism in gauge systems with first class constraints \cite{CERNrep,BV,BFV,Henneaux}. We specify it to the case of the time parametrization invariant gravity theory of spatially {\em closed} cosmology with a total Hamiltonian which reduces to the linear combination of Hamiltonian and momentum constraints \cite{BarvU}. Within the BRST technique of the relativistic phase space of gauge and ghost fields \cite{CERNrep} we build the operator of unitary evolution in their representation space. In gravity theory of the above type it is the operator of evolving BRS transformation in the representation space of the theory, which later in Sect.4 will be used to construct the projector (\ref{rhocanonical}). In Sect.3 we describe a special Batalin-Marnelius (BM) procedure of gauging out this symmetry \cite{BM,FHP}, which allows one to construct a well-defined physical inner product in the space of BRS singlets, circumventing a well-known problem of its $0\times\infty$-indeterminacy for BRS invariant states \cite{Henneaux,BarvU}. In Sect.4 we perform a truncation of the BFV formalism to the Dirac quantization scheme \cite{BKr,BarvU}, which serves as a realization of the BM gauge fixing and build a projector (\ref{rhocanonical}) onto the subspace of Dirac constraints. Here we specify the integration measure in the path integral (\ref{rhocanonical}) and prove basic properties of the latter -- annihilation of the kernel (\ref{rhocanonical}) under the action of Dirac constraints on its arguments, gauge independence and independence of the choice of the ``time" segment $[t_-,t_+]$. Finally we derive the expression (\ref{Z}) for the microcanonical statistical sum by tracing the projector (\ref{rhocanonical}) with respect to the physical inner product. Concluding section summarizes the results and gives a brief overview of open issues in BRST formalism for the cosmological density matrix.

\section{BFV formalism}
Diffeomorphism invariant gravity theory with the Einstein-Hilbert action of the metric and matter fields in the canonical formalism has the ADM form and incorporates  splitting the full configuration space of $g_{\mu\nu},\phi$ into phase space coordinates $q^i=g_{ab}({\bf x}),\phi({\bf x})$, $a=1,2,3$, and non-dynamical Lagrange multipliers $N^\mu=(N^\perp({\bf x}), N^a({\bf x}))$
        \begin{eqnarray}
        S=\int dt\,\Big\{\,p_i\dot{q}^i-H_0(q,p)
        -N^\mu H_\mu(q,p)\Big\}.           \label{1.1}
        \end{eqnarray}
Here $p_i$ are the momenta conjugated to $q^i$, whereas the variables $N^\mu$ do not have conjugated momenta. Like in Introduction we will use condensed canonical notations when the field labels carry together with discrete indices also spatial coordinates, and contraction of indices implies spatial (but not time) integration. The variation of $N^\mu$ leads to the set of nondynamical equations -- the Hamiltonian and momentum constraints on phase space variables
        \begin{eqnarray}
        H_\mu(q,p)=0.                            \label{1.2}
        \end{eqnarray}
In open models with asymptotically flat (or other boundaries like horisons, etc.) $H_0(q,p)$ represents the relevant surface integral specified by boundary conditions which are the part of physical setting for the gravitational system. Below we consider the case of spatially closed cosmology with $H_0(q,p)=0$.

The diffeomorphism invariance of the theory has a manifestation in Poisson bracket algebra of constraint functions $H_\mu(q,p)$
        \begin{eqnarray}
        &&\{H_\mu,H_\nu\}=U^\lambda_{\mu\nu}
        H_\lambda,                             \label{1.3}
        \end{eqnarray}
with the structure functions $U^\lambda_{\mu\nu}=U^\lambda_{\mu\nu}(q)$ which depend on phase-space coordinates $q$. Thus already at the classical level these constraints form an {\em open} group algebra which under quantization should go over to the commutator algebra of constraint operators $\hat H_\mu$ acting in the representation space of canonically commuting $\hat q^i,\hat p_i$,
        \begin{eqnarray}
        [\hat{H}_\mu,\hat{H}_\nu]=
        i\hat{U}^{\lambda}_{\mu\nu}
        \hat{H}_\lambda.                \label{comalgebra}
        \end{eqnarray}
At the quantum level the classical constraints take the form of equations on physical quantum states $|\,\varPsi\,\rangle$
        \begin{eqnarray}
        \hat H_\mu |\,\varPsi\,\rangle=0.   \label{Dirac}
        \end{eqnarray}
In the functional coordinate representation of quantum gravity they form the system of the Wheeler-DeWitt equations on the wavefunction $\langle\,q\,|\,\varPsi\,\rangle=\varPsi(q)$.

Their consistency requires the commutator algebra (\ref{comalgebra}) to hold with the operator structure functions $\hat{U}^{\lambda}_{\mu\nu}$ standing to the left of $\hat{H}_\lambda$. The operator ($q$-number) nature of $\hat{U}^{\lambda}_{\mu\nu}$ implies that the algebra of quantum constraints is also open, and this precludes one from a simple construction of the projector (\ref{projector}), which for example is available by the procedure of group integration for closed compact Lie groups \cite{MarneliusGroup}. To circumvent this difficulty we will construct this projector from the unitary evolution operator in the BRST quantization approach.

This approach generalizes the Dirac quantization scheme (\ref{comalgebra})-(\ref{Dirac}) by extending the representation space of the original (usually called ``matter") phase space variables $q^i,p_i$ to that of the relativistic phase space variables. The latter include together with $q^i,p_i$ the canonically conjugated pairs of Lagrange multipliers and their momenta $\pi_\mu$ and canonical pairs of Grassman (fermionic) ghosts ${C^\mu, \cal P_\mu}$ and anti-ghosts $\bar C_\mu, \bar{\cal P}^\mu$,
        \begin{eqnarray}
        &&Q^I,P_I=q^i,p_i;\,N^\mu,\pi_\mu;\,
        C^\mu,  {\cal P}_\mu;\,
        \bar C_\mu, \bar {\cal P}^\mu,    \label{pairs}\\
        &&[Q^I,P_J]=i\,\delta^I_J.  \label{cancom}
        \end{eqnarray}
Here $[A,B]$ denotes a supercommutator taking into account the Grassman pairity $n$ of $A$ and $B$
        \begin{eqnarray}
        [A,B]_\pm=AB-(-1)^{n(A) n(B)}BA.
        \end{eqnarray}
In gravity theory with bosonic matter fields $n(q)=n(N)=0$ and $n(C)=n({\cal P})=n(\bar C)=n(\bar{\cal P})=1$. The canonical commutation relations (\ref{cancom}) represent the quantum version of classical Poisson superbracket commutators $\{Q^I,P_J\}=i\,\delta^I_J$ (we use units with $\hbar=1$). Ghost variables have Hermiticity properties compatible with these commutation relations
        \begin{eqnarray}
        C^{\mu\dagger}=C^\mu,\quad
        {\cal P}_\mu^\dagger=-{\cal P}_\mu,\quad
        \bar C_\mu^\dagger=-\bar C_\mu,\quad
        \bar{\cal P}^{\mu\dagger}
        =\bar{\cal P}^\mu.          \label{Hermiticity}
        \end{eqnarray}

In what follows we will regularly omit the hat notation for operators (\ref{pairs}) and only use it in case when they have to be distinguished from their $c$-number eigenvalues. The hat notations will as a rule be used for composite operators. State vectors in the representation space of all relativitic variables will be denoted by double ket notations $||\,{\mbox{\boldmath$\varPsi$}}\rangle\rangle$ (contrary to vectors $|\,\varPsi\,\rangle$ in the representation space of ``matter" operators $(\hat q^i,\hat p_i)$). The coordinate representation will be introduced as follows,
        \begin{eqnarray}
        &&||\,Q\,\rangle\rangle\equiv||\, q,N,C,\bar C\,\rangle\rangle, \quad \hat Q^I||\,Q\,\rangle\rangle=Q^I||\,Q\,\rangle\rangle,\\
        &&{\mbox{\boldmath$\varPsi$}}(Q)=\langle\langle\,Q\,||\, {\mbox{\boldmath$\varPsi$}}\,\rangle\rangle,\\
        &&\langle\langle\, {\mbox{\boldmath$\varPsi$}}_1\,||
        \,{\mbox{\boldmath$\varPsi$}}_2\rangle\rangle=\int dQ\,{\mbox{\boldmath$\varPsi$}}_1
        \vphantom{1}^*(Q)\,
        {\mbox{\boldmath$\varPsi$}}_2(Q),    \label{BRSprod}
        \end{eqnarray}
where the BRS inner product is defined in the sense of Berezin integration over $Q$. To finish description of notations for BFV formalism we mention that we will also need the momentum representation in the sector of the Lagrangian multipliers with the interchanged roles of $N^\mu$ and $\pi_\mu$. It will be denoted by tilde, and the corresponding set of variables will look like
        \begin{eqnarray}
        &&\tilde Q^I,\tilde P_I=q^i,p_i;\,\pi_\mu,-N^\mu;\,
        C^\mu,  {\cal P}_\mu;\,
        \bar C_\mu, \bar {\cal P}^\mu,            \label{momrep}\\
        &&\tilde{\mbox{\boldmath$\varPsi$}}(\tilde Q)=\langle\langle\,\tilde Q\,||\, {\mbox{\boldmath$\varPsi$}}\,\rangle\rangle
        \end{eqnarray}

The basic object of the BRST/BFV technique is the nilpotent fermionic BRS operator $\hat\varOmega$ acting in the space of $||\,{\mbox{\boldmath$\varPsi$}}\rangle\rangle$ and satisfying the master equation
        \begin{eqnarray}
        &&[\hat\varOmega,\hat\varOmega]
        \equiv\hat\varOmega^2=0.
        \end{eqnarray}
This equation allows one to look for the solution as an expansion in powers of the ghosts $C^\mu$ and their momenta ${\cal P}_\mu$ starting with the combination $\hat\varOmega=\pi_\alpha\bar{\cal P}^\alpha+C^\mu \hat H_\mu+O({\cal P}C^2)$. The coefficients of this expansion $\hat H_\mu,\hat U^\lambda_{\mu\nu},\hat U^{\lambda\sigma}_{\mu\nu\alpha},...$ are the structure functions of the gauge algebra of constraints beginning with (\ref{comalgebra}) -- higher order structure functions follow from applying the Jacobi identity to multiple commutators of (\ref{comalgebra}) with $\hat H_\sigma$. In non-supersymmetric gravity theory, which we consider here, this sequence terminates at $\hat U^{\lambda\sigma}_{\mu\nu\alpha} =0$, and $\hat\varOmega$ takes the form
        \begin{eqnarray}
        \hat\varOmega=\pi_\alpha\bar{\cal P}^\alpha+C^\mu \hat H_\mu+\frac12\,C^\nu C^\mu \hat U^\lambda_{\mu\nu}{\cal P}_\lambda, \quad
        \hat\varOmega^\dagger=\hat\varOmega.
        \end{eqnarray}
It is Hermitian in the BRS inner product (\ref{BRSprod}) in accordance with the Hermiticity properties of ghost variables (\ref{Hermiticity}), provided the quantum Dirac constraints have the anti-Hermitian part\footnote{In higher rank gauge theories with nonvanishing higher order structure functions this Hermiticity properties are modified by their higher order contributions.}
        \begin{eqnarray}
        \hat H_\mu-\hat H_\mu^\dagger=
        i\hat{U}^\lambda_{\mu\lambda}.          \label{2.9}
        \end{eqnarray}

In models with $H_0(q,p)\neq 0$ this BRS operator determines also its BRS extension $\hat{\cal H}=\hat H_0+O(C{\cal P})$ by the equation $[\hat\varOmega,{\cal\hat H}_0]=0$ and the so-called {\em unitarizing} Hamiltonian
        \begin{eqnarray}
        \hat{\cal H}_{\varPhi}= \hat{\cal H}_0+\frac1{i}[\,\hat\varPhi,\hat\varOmega\,].
        \end{eqnarray}
It explicitly depends on the gauge fermion $\varPhi$, $n(\hat\varPhi)=1$, which provides gauge fixing in the BRST/BFV formalism. In parametrization invariant closed cosmology $\hat{\cal H}_0=0$, and the unitarizing Hamiltonian reduces to the commutator of the BRS operator and gauge fermion.

The unitary evolution operator $\hat{\mbox{\boldmath$U$}}_{\varPhi}(t,t_-)$ acting in the space of $||\,{\mbox{\boldmath$\varPsi$}}\rangle\rangle$ is a solution of the following Cauchy problem
        \begin{eqnarray}
        &&i\hbar\frac{\partial}{\partial t}\hat{\mbox{\boldmath$U$}}_{\varPhi}(t,t_-)=
        \hat{\cal H}_{\varPhi}
        \hat{\mbox{\boldmath$U$}}
        _{\varPhi}(t,t_-),            \label{Schroedinger}\\
        &&\hat{\mbox{\boldmath$U$}}
        _{\varPhi}(t_-,t_-)
        =\mathbb{I}
        \end{eqnarray}
It is obvious that from $[\hat\varOmega,[\hat\varPhi,\hat\varOmega]]\equiv 0$ and $[\hat\varOmega,\hat{\cal H}_{\varPhi}]=0$ the BRS operator is a constant of motion in this evolution,
        \begin{eqnarray}
        [\,\hat\varOmega,
        \hat{\mbox{\boldmath$U$}}
        _{\varPhi}(t,t_-)\,]=0,    \label{com}
        \end{eqnarray}
so that it plays the role of the conserved BRS charge and serves as a generator of BRS transformations in the relativistic phase space.

In the coordinate representation the kernel of the unitary evolution has a representation of the canonical path integral
        \begin{eqnarray}
        &&{\mbox{\boldmath$U$}}_{\varPhi}(\,t_+,Q_+|\,t_-,Q_-)
        \equiv\langle\langle\, Q_+\,||\,\hat{\mbox{\boldmath$U$}}
        _{\varPhi}(t_+,t_-)\,||\,Q_-\,\rangle\rangle\nonumber\\
        &&\qquad\qquad\qquad=\int\limits_{Q(t_\pm)=Q_\pm} D[\,Q,P\,]\,\exp\left\{i\int_{t_-}^{t_+} dt\,\left(P_I\dot Q^I
        -{\cal H}_{\varPhi}(Q,P)
        \right)\right\},                \label{unitarykernel}
        \end{eqnarray}
where ${\cal H}_{\varPhi}(Q,P)$ is the $QP$-symbol of the unitarizing Hamiltonian given by the Poisson superbracket of $c$-number symbols $\varPhi$ and $\varOmega$ of the operator gauge fermion and BRS charge
        \begin{eqnarray}
        {\cal H}_{\varPhi}(Q,P)=\{\,\varPhi,\varOmega\,\}
        \end{eqnarray}
(remember that $H_0=0$). Also, $D[\,Q,P\,]$ is a Liouville integration measure in the full boson-fermion phase space of $c$-number histories
        \begin{eqnarray}
        D[\,Q,P\,]= \prod\limits_t dQ(t)\,\prod\limits_{t^*}dP(t^*).
        \end{eqnarray}
The difference between the set of points $t=(t_N,...t_1)$ and $t^*=(t^*_{N+1},...t^*_1)$, $N\to\infty$, over which the product of integration measure factors is taken, reflects the typical slicing of the path integral into a sequence of multiple integrals in the decomposition of the full time segment $[t_+,t_-]$ into infinitesimal pieces. This decomposition, $t_+>t_N>t_{N-1}>...>t_1>t_-$, $t_+>t^*_{N+1}>t_N>t^*_N>...>t_1>t^*_1>t_-$, implies that the points $t^*$, at which the integrated momenta are taken, are associated with ``interiors" of the segments $[t_{i+1},t_i]$ whose boundaries carry the integrated coordinates -- so that the number of momentum integrations is by one larger than those of coordinate ones.

In the momentum representation for Lagrangian multipliers $\tilde Q^I=q^i,\pi^\mu,C^\mu,\bar C_\mu$, cf. (\ref{momrep}), the unitary evolution kernel has a similar path integral representation
        \begin{eqnarray}
        &&\tilde{\mbox{\boldmath$U$}}_{\varPhi}(\,t_+,\tilde Q_+|\,t_-,\tilde Q_-)
        \equiv\langle\langle\, \tilde Q_+\,||\,\hat{\mbox{\boldmath$U$}}
        _{\varPhi}(t_+,t_-)\,||\,\tilde Q_-\,\rangle\rangle\nonumber\\
        &&\qquad\qquad\qquad=\int\limits_{\tilde Q(t_\pm)=\tilde Q_\pm} \tilde D[\,Q,P\,]\,
        \exp\left\{i\int_{t_-}^{t_+} dt\,
        \left(\tilde P_I\dot {\tilde Q}^I
        -{\cal H}_{\varPhi}(Q,P)\right)
        \right\}                      \label{unitarykernel1}\\
        &&\tilde D[\,Q,P\,]=
        \prod\limits_t d\tilde Q(t)\,
        \prod\limits_{t^*}d\tilde P(t^*)
        \end{eqnarray}
and is of course related by the Fourier transform to the kernel (\ref{unitarykernel})
        \begin{eqnarray}
        &&\tilde{\mbox{\boldmath$U$}}_{\varPhi}(\,t_+,\tilde Q_+|\,t_-,\tilde Q_-)
        =\int dN_+dN_-\,e^{-i\pi_+N_+}{\mbox{\boldmath$U$}}
        _{\varPhi}(\,t_+,Q_+|\,t_-,Q_-)\,
        e^{i\pi_-N_-}         \label{unitarykernel2}
        \end{eqnarray}
in full accordance with the fact that two symplectic forms in the integrands of path integrals on the left and right hand sides here are related by
        \begin{eqnarray}
        \int_{t_-}^{t_+} dt\,\tilde P\dot {\tilde Q}=
        \int_{t_-}^{t_+} dt\,P\dot Q
        -\pi_+N_++\pi_-N_-, \quad \pi_\pm\equiv\pi(t_\pm),\quad N_\pm\equiv N(t_\pm).
        \end{eqnarray}

The principal theorem of the BFV quantization is that the matrix elements of the unitary evolution operator $\hat{\mbox{\boldmath$U$}}_{\varPhi}(t,t_-)$ between the BRS-invariant physical states annihilated by $\hat\varOmega$ are independent of the choice of the gauge fermion
        \begin{eqnarray}
        \hat\varOmega\,||
        \,{\mbox{\boldmath$\varPsi$}}_{\!1,2}
        \rangle\rangle=0\quad
        \Rightarrow\quad
        \delta_{\varPhi}\langle\langle\, {\mbox{\boldmath$\varPsi$}}_{\!1}||
        \,\hat{\mbox{\boldmath$U$}}
        _{\varPhi}(t_+,t_-)\,||
        \,{\mbox{\boldmath$\varPsi$}}_2
        \rangle\rangle=0.                   \label{theorem}
        \end{eqnarray}

The logic of the above BRST/BFV construction is based on the observation that relativistic gauge conditions, involving time derivatives of Lagrange multipliers, make the latter propagating and having nonvanishing canonical momenta $\pi_\mu$ which are absent in the original action (\ref{1.1}). To compensate the contribution of these artificially introduced degrees of freedom and the degrees of freedom which have to be excluded by first class constraints (\ref{1.2}) one introduces dynamical ghosts and antighosts of the statistics opposite to those of $\hat H_\mu$. Due to statistics they effectively subtract in quantum loops the contribution of these gauge degrees of freedom. However, a similar subtraction should be done in external lines of Feynman diagrams, which means that not all quantum states in BRST space are physical. Physical states form a subspace belonging to the kernel of the BRS operator. In this subspace due to the above theorem the transition amplitudes and quantum averages are independent of the choice of gauge fixing procedure -- the corner stone of quantizing the gauge invariant systems.

\section{Batalin-Marnelius gauge fixing and the physical inner product}

Physical states should be BRS invariant
        \begin{eqnarray}
        \hat\varOmega\,||\,{\mbox{\boldmath$\varPsi$}}\,
        \rangle\rangle=0.
        \end{eqnarray}
This equation does not uniquely select its solution because in view of the nilpotent nature of $\hat\varOmega$ the BRS transformed state $||\,{\mbox{\boldmath$\varPsi$}}\,\rangle\rangle'$,
        \begin{eqnarray}
        ||\,{\mbox{\boldmath$\varPsi$}}\,\rangle\rangle'
        =||\,{\mbox{\boldmath$\varPsi$}}\,\rangle\rangle+
        \hat\varOmega\,||\,{\mbox{\boldmath$\varPhi$}}\,
        \rangle\rangle,               \label{BRSsym}
        \end{eqnarray}
with an arbitrary $||\,{\mbox{\boldmath$\varPhi$}}\, \rangle\rangle$ also satisfies the BRS equation. This invariance results in the problem of constructing (or regulating) the physical inner product. Problem is that the original inner product (\ref{BRSprod}) for physical states represents the $0\times\infty$-indeterminacy. The essence of this indeterminacy can be qualitatively explained by the fact that squaring of the physical state ${\mbox{\boldmath$\varPsi$}}(Q)\sim\delta(\hat\varOmega)$ in (\ref{BRSprod}) gives a divergent factor whereas the integration over Grassman variables multiplies it by zero. This inner product can be regulated by transforming the BRS-invariant state $||\,{\mbox{\boldmath$\varPhi$}}\, \rangle\rangle$ to a special gauge as it was suggested by Batalin and Marnelius in \cite{BM} (see also \cite{FHP})
        \begin{eqnarray}
        ||\,{\mbox{\boldmath$\varPsi$}}\,\rangle\rangle\to
        ||\,{\mbox{\boldmath$\varPsi$}}_{B\!M}\rangle\rangle:\quad
        \hat{\cal P}_\mu||\,{\mbox{\boldmath$\varPsi$}}_{B\!M}\rangle\rangle=0,\;
        \hat N^\mu||\,{\mbox{\boldmath$\varPsi$}}_{B\!M}
        \rangle\rangle=0,\;
        {\hat{\bar{\cal P}}^\mu} ||\,{\mbox{\boldmath$\varPsi$}}_{B\!M}
        \rangle\rangle=0                   \label{BMgfixing}
        \end{eqnarray}
(the third condition is in fact a corollary of the second one, $[\hat\varOmega,N^\mu]\,||\,{\mbox{\boldmath$\varPsi$}}_{B\!M} \rangle\rangle=0$). The consistency of this gauge with the BRS-invariance of $||\,{\mbox{\boldmath$\varPsi$}}_{B\!M} \rangle\rangle$ implies that it also satisfies the quantum Dirac constraints
        \begin{eqnarray}
        &&0=\frac1i\,[\,\hat\varOmega,\hat{\cal P}_\mu]\,||\,{\mbox{\boldmath$\varPsi$}}_{B\!M}\rangle\rangle
        =(\hat H_\mu+C^\nu \hat U^\lambda_{\nu\mu}\hat{\cal P}_\lambda)\, ||\,{\mbox{\boldmath$\varPsi$}}_{B\!M}\rangle\rangle
        =\hat H_\mu\, ||\,{\mbox{\boldmath$\varPsi$}}_{B\!M}\rangle\rangle,
        \end{eqnarray}
and its wavefunction is independent of ghost variables,
        \begin{eqnarray}
        &&\frac\partial{\partial C^\mu}{\mbox{\boldmath$\varPsi$}}_{B\!M}(Q)=0,\quad
        \frac\partial{\partial \bar C_\mu}{\mbox{\boldmath$\varPsi$}}_{B\!M}(Q)=0
        \end{eqnarray}
Together with (\ref{BMgfixing}) this means that ${\mbox{\boldmath$\varPsi$}}_{B\!M}(Q)$ has the form
        \begin{eqnarray}
        &&\langle\langle\,Q\,||\,
        {\mbox{\boldmath$\varPsi$}}_{B\!M}\rangle\rangle
        =\langle\,q\,|\,\varPsi\,
        \rangle\,\delta(N)\equiv
        \varPsi(q)\,\delta(N),             \label{BMwf}
        \end{eqnarray}
where the ``matter" part $\varPsi(q)$ satisfies quantum Dirac constraints in the coordinate representation,
        \begin{eqnarray}
        &&\hat H_\mu\varPsi(q)=0.
        \end{eqnarray}

According to \cite{BM} the physical inner product of wavefunctions $||\,
{\mbox{\boldmath$\varPsi$}}_{B\!M}\rangle\rangle$ can be regularized by a special operator-valued measure which is explicitly built with the aid of the gauge fermion $\hat\varPhi$
        \begin{eqnarray}
        &&\langle\langle\, {\mbox{\boldmath$\varPsi$}}'\,||
        \,{\mbox{\boldmath$\varPsi$}}\rangle\rangle_{\rm phys}=
        \langle\langle\, {\mbox{\boldmath$\varPsi$}}'_{B\!M}||
        \,e^{[\,\hat\varPhi,\hat\varOmega\,]}||
        \,{\mbox{\boldmath$\varPsi$}}_{B\!M}
        \rangle\rangle,                      \label{innerprod}
        \end{eqnarray}
The choice of this fermion is immaterial because
        \begin{eqnarray}
        &&\delta_{\varPhi}\langle\langle\, {\mbox{\boldmath$\varPsi$}}'\,||
        \,{\mbox{\boldmath$\varPsi$}}\rangle\rangle_{\rm phys}=\int_0^1 ds\,
        \langle\langle\, \tilde{\mbox{\boldmath$\varPsi$}}'_{B\!M}||\,
        e^{(1-s)[\,\hat\varPhi,\hat\varOmega\,]}\,
        [\,\delta\hat\varPhi,\hat\varOmega\,]\,
        e^{s[\,\hat\varPhi,\hat\varOmega\,]}||
        \,{\mbox{\boldmath$\varPsi$}}_{B\!M}\rangle\rangle=0
        \end{eqnarray}
since $[\hat\varOmega,[\hat\varPhi,\hat\varOmega]]=0$ and $\hat\varOmega\,||
\,{\mbox{\boldmath$\varPsi$}}_{B\!M}\rangle\rangle=0$. However, with a special choice of this fermion the physical inner product (\ref{innerprod}) resolves the $0\times\infty$ uncertainty and becomes well defined \cite{BM}.\footnote{In \cite{BM} the factor $\exp{[\,\hat\varPhi,\hat\varOmega\,]}$ was introduced as the operator coefficient relating $||\,
{\mbox{\boldmath$\varPsi$}}\,\rangle\rangle$ and $||\,
{\mbox{\boldmath$\varPsi$}}_{B\!M}\rangle\rangle$ rather than the physical inner product measure. The treatment of this factor as this measure, which we adopt here, is essentially equivalent to the presentation of \cite{BM}.}

This is easy to show if this fermion is constructed with the aid of "matter" gauge conditions functions $\hat\chi^\mu=\chi^\mu(\hat q,\hat p)$ which commute with themselves, $[\hat\chi^\mu,\hat\chi^\nu]=0$, and with the structure functions operators, $[\hat\chi^\mu,\hat U^\lambda_{\alpha\beta}]=0$ (in gravity theory this is coordinate gauge conditions $\chi^\mu(q)$ commuting with $U^\lambda_{\alpha\beta}=U^\lambda_{\alpha\beta}(q)$). For the fermion $\hat\varPhi_{B\!M}$ in the form
        \begin{eqnarray}
        &&\hat\varPhi_{B\!M}=
        \hat{\bar C}_\mu\hat\chi^\mu,     \label{BMferm}\\
        &&[\hat\varPhi_{B\!M},\hat\varOmega]
        =i\pi_\mu\hat\chi^\mu
        +i\bar C_\mu\hat J^\mu_\nu C^\nu,  \label{comPhiQ}\\
        &&\hat J^\mu_\nu=\frac1i\,
        [\,\hat\chi^\mu,\hat H_\nu],    \label{BRSJ}
        \end{eqnarray}
we have the physical inner product in the coordinate representation with $\hat\pi_\mu=\partial/i\partial N^\mu$
        \begin{eqnarray}
        &&\langle\langle\, {\mbox{\boldmath$\varPsi$}}'\,||
        \,{\mbox{\boldmath$\varPsi$}}\rangle\rangle_{\rm phys}=\int dq\,dN\,dC\,d\bar C\, \varPsi'^*(q)\,\delta(N)\,\exp\left(-i\bar C_\mu\hat J^\mu_\nu C^\nu+\hat\chi^\mu\frac\partial{\partial N^\mu}\right)\,\delta(N)\,\varPsi(q)\nonumber\\
        &&\qquad\qquad\qquad=\langle\,\varPsi'\,|\int d\pi\,dC\,d\bar C\,
        e^{-i\bar C_\mu\hat J^\mu_\nu C^\nu+i\pi_\mu\hat\chi^\mu}
        |\,\varPsi\,\rangle.           \label{physprod}
        \end{eqnarray}
Here we took into account a special form of Batalin-Marnelius wavefunctions (\ref{BMwf}). Their special dependence on Lagrange multipliers and ghost variables leads to the construction of the physical inner product as a special operator valued measure in the space of Dirac wavefunctions $|\,\varPsi'\,\rangle$ and $|\,\varPsi\,\rangle$. This measure is known in quadratures as an explicit integral over ghost fields and Lagrangian multipliers momenta,
        \begin{eqnarray}
        \hat M=\int d\pi\,dC\,d\bar C\,
        e^{-i\bar C_\mu\hat J^\mu_\nu C^\nu+i\pi_\mu\hat\chi^\mu}
        =\delta(\hat\chi)\,
        \det\hat J^\mu_\nu\,\Big(\,1+
        O\big(\,[\,\hat\chi,\hat J\,]\,
        \big)\Big).                         \label{measure}
        \end{eqnarray}
This representation allows one to disentangle in the leading order the delta function of gauge conditions, $\delta(\hat\chi)=\prod_\mu\delta(\hat\chi^\mu)$, which is well defined in view of their commutativity.\footnote{This, however, does not save us from extra corrections, because $J^\mu_\nu(\hat q,\hat p)$ depends on the momentum $\hat p_i$ and is a differential operator acting in the space of $q$.} The mechanism of regulating the $0\times\infty$ uncertainty by the BRS measure (\ref{innerprod}) is obvious from Eqs. (\ref{physprod}) and (\ref{measure}). The $0$-factor of grassman integration is regulated by the Gaussian integrand in (\ref{physprod}), divergent factor $\sim[\delta(N)]^2$ is regulated due to the shift operator in (\ref{physprod}) acting in the space of $N^\mu$ and, finally, the divergent factor bilinear in $\varPsi'^*(q)$ and $\varPsi(q)$, $\varPsi'^*(q)\varPsi(q)\sim[\delta(\hat H)]^2$, is regulated by the delta function of gauge conditions $\delta(\chi)$.

The expression (\ref{measure}) is the analogue of the time-local measure in the canonical Faddeev-Popov path integral \cite{Faddeev} with the operator (\ref{BRSJ}) semiclassically represented by the Poisson bracket $\hat J^\mu_\nu=\{\chi^\mu,H_\nu\}$. In the semiclassical approximation (one-loop order) it was suggested in \cite{GenSem,BKr,BarvU,geom},
        \begin{eqnarray}
        &&\langle\langle\, {\mbox{\boldmath$\varPsi$}}'\,||
        \,{\mbox{\boldmath$\varPsi$}}\rangle\rangle_{\rm phys}=\int dq\, \varPsi'^*(q)\,
        \delta(\chi(q))\,\det\hat J^\mu_\nu\,\,\varPsi(q)+O(\hbar). \label{innerprod1}
        \end{eqnarray}
and explicitly shown to be independent of the choice of gauge conditions $\chi^\mu(q)$ for semiclassical solutions of quantum Dirac constraints $\varPsi(q)$ and $\varPsi'(q)$. This nontrivial operator measure also guarantees the Hermiticity property (\ref{2.9}), which confirms the consistency of the whole formalism \cite{geom}.

\section{Projector on the space of quantum Dirac constraints}

Transition to the BM gauge can in fact be obtained by a simple procedure of truncation of BRS invariant wavefunctions to the sector of ``matter" variables, that was suggested in \cite{BKr,BarvU}. Introduce a wavefunction $\varPsi(q)$ in the representation space of $\hat q^i,\hat p_i$ which can be obtained from the solution $||\,{\mbox{\boldmath$\varPsi$}}\rangle\rangle$ of the BRS equation $\hat\varOmega||\,{\mbox{\boldmath$\varPsi$}}\rangle\rangle=0$ by
        \begin{eqnarray}
        \varPsi(q)=\int dN\,
        {\mbox{\boldmath$\varPsi$}}
        (q,N,C,\bar C)\,\big|_{\,C=0}.  \label{truncation}
        \end{eqnarray}
As we show below, from the BRS equation it follows that this function satisfies quantum Dirac constraints and is independent of the antighost variable $\bar C$ (that is why the argument $\bar C$ of the right hand side is omitted on the left hand side of this definition)
        \begin{eqnarray}
        \hat H_\mu\varPsi(q)=0, \quad \frac\partial{\partial\bar C_\mu}
        \varPsi(q)=0.           \label{Dirac}
        \end{eqnarray}
Both properties follow from the BRS equation which in the coordinate representation reads as
        \begin{eqnarray}
        &&\hat\varOmega\,{\mbox{\boldmath$\varPsi$}}(Q)
        =\left(\frac\partial{\partial N^\mu}\frac\partial{\partial \bar C_\mu}+C^\mu\hat H_\mu+\frac{i}2\, C^\nu C^\mu \hat U^\lambda_{\mu\nu}\frac\partial{\partial C^\lambda}\right)\,
        {\mbox{\boldmath$\varPsi$}}(Q)=0.   \label{BRSeq}
        \end{eqnarray}
Integrating this equation over the Lagrange multipliers $N$ in infinite limits and assuming that ${\mbox{\boldmath$\varPsi$}}(Q)$ falls off sufficiently rapidly at $N\to\pm\infty$, one finds that the first term of (\ref{BRSeq}) vanishes. Then one can differentiate the result with respect to the ghost field $C^\mu$ and subsequently put $C=0$. In operator notations this sequence of operations means
        \begin{eqnarray}
        0=\frac1i\hat{\cal P}_\mu\int dN\,\hat\varOmega\,
        {\mbox{\boldmath$\varPsi$}}(Q)\,\big|_{\,C=0}
        =\hat H_\mu\varPsi(q)
        \end{eqnarray}
and proves the first of relations (\ref{Dirac}). The second relation follows from multiplying Eq. (\ref{BRSeq}) by $N^\mu$, putting $C=0$ and integrating over $N$ by parts in the remaining first term of this equation
        \begin{eqnarray}
        0=\int dN\,N^\mu\,\hat\varOmega\,
        {\mbox{\boldmath$\varPsi$}}(Q)\,\big|_{\,C=0}
        =\int dN\,N^\mu\,\frac\partial{\partial N^\nu}\frac\partial{\partial \bar C_\nu}
        {\mbox{\boldmath$\varPsi$}}(Q)\,\big|_{\,C=0}=
        -\frac\partial{\partial\bar C_\mu}
        \varPsi(q).
        \end{eqnarray}

This truncation of the BRST quantization scheme to the Dirac quantization suggested in \cite{BKr,BarvU} serves in fact as a realization of the Batalin-Marnelius gauge fixing (\ref{BMgfixing}) of the BRST symmetry (\ref{BRSsym}). To put the generic BRS state into the Batalin-Marnelius gauge it is enough to take the bosonic ``body" of its wavefunction, integrate it over the Lagrange multipliers argument $N$ and multiply by $\delta(N)$,
        \begin{eqnarray}
        ||\,{\mbox{\boldmath$\varPsi$}}\rangle\rangle\to
        ||\,{\mbox{\boldmath$\varPsi$}}_{B\!M}\rangle\rangle:\quad
        {\mbox{\boldmath$\varPsi$}}_{B\!M}(Q)=
        \delta(N)\int dN'\,{\mbox{\boldmath$\varPsi$}}(q,N',C,\bar C)\,\big|_{\,C=0}.
        \end{eqnarray}

Truncation similar to (\ref{truncation}) for the kernel of the unitary evolution (\ref{unitarykernel}) reads
        \begin{eqnarray}
        &&U(q_+,q_-)=\int dN_+\,dN_-\,
        {\mbox{\boldmath$U$}}_{\varPhi}
        (\,t_+,Q_+|\,t_-,Q_-)\,\big|_{C_\pm=0}.   \label{U}
        \end{eqnarray}
This object can be represented as a matrix element of $\hat{\mbox{\boldmath$U$}}_{\varPhi}(t_+,t_-)$ between the following two states $||\,{\mbox{\boldmath$\varPsi$}}_\pm\rangle\rangle$ which are both zero vectors of the Lagrangian multiplier momentum and trivially satisfy the BRS equation,
        \begin{eqnarray}
        &&U(q_+,q_-)\equiv\langle\langle\, {\mbox{\boldmath$\varPsi$}}_+||
        \,\hat{\mbox{\boldmath$U$}}_{\varPhi}(t_+,t_-)\,||
        \,{\mbox{\boldmath$\varPsi$}}_-\rangle\rangle,\\
        &&{\mbox{\boldmath$\varPsi$}}_\pm(q,N,C,\bar C)=\delta(q-q_\pm)\delta(C)\delta(\bar C),\quad \hat\pi_\alpha||\,{\mbox{\boldmath$\varPsi$}}_\pm\rangle\rangle=0,
        \quad
        \hat\varOmega\,
        ||\,{\mbox{\boldmath$\varPsi$}}_\pm\rangle\rangle=0.
        \end{eqnarray}
Therefore, in virtue of the main theorem of BRS quantization it is independent of the choice of the gauge fermion $\varPhi$ in $\hat{\mbox{\boldmath$U$}}_{\varPhi}(t_+,t_-)$
        \begin{eqnarray}
        &&\delta_\varPhi U(q_+,q_-)=0,   \label{gaugeind}
        \end{eqnarray}
which guarantees the uniqueness of its definition. The second important property is that the kernel (\ref{U}) is independent of $t_\pm$ in parametrization invariant theory with $H_0=0$ in (\ref{1.1}) because
        \begin{eqnarray}
        i\frac\partial{\partial t_+}U(q_+,q_-)=\langle\langle\, {\mbox{\boldmath$\varPsi$}}_+||\,
        \frac1{i}\,[\,\hat\varPhi,\hat\varOmega\,]
        \,\hat{\mbox{\boldmath$U$}}_{\varPhi}(t_+,t_-)\,||
        \,{\mbox{\boldmath$\varPsi$}}_-
        \rangle\rangle=0         \label{timeind}
        \end{eqnarray}
in view of the Schroedinger equation (\ref{Schroedinger}) for $\hat{\mbox{\boldmath$U$}}_{\varPhi}(t_+,t_-)$. This allows one to omit $\varPhi$ and $t_\pm$ labels in the left hand side of the definition (\ref{U}).

Finally, applying the same as above derivations to the main BRS equation (\ref{com}) for $\hat{\mbox{\boldmath$U$}}_{\varPhi}(t_+,t_-)$, i.e.
        \begin{eqnarray}
        &&0=\int dN_+ dN_-N^\mu_+\langle\langle\,Q_+||\,[\,\hat\varOmega,
        \hat{\mbox{\boldmath$U$}}
        _{\varPhi}(t_+,t_-)\,]\,||
        \,Q_-\rangle\rangle\,\Big|_{\,C_\pm=0}=
        -\frac\partial{\partial\bar C^+_\mu}U(q_+,q_-),\\
        &&0=\frac\partial{\partial C^\mu_+}\int dN_+ dN_-\langle\langle\,Q_+||\,[\,\hat\varOmega,
        \hat{\mbox{\boldmath$U$}}
        _{\varPhi}(t_+,t_-)\,]\,||
        \,Q_-\rangle\rangle\,
        \Big|_{\,C_\pm=0}=
        \hat H_\mu\,U(q_+,q_-),    \label{projU}
        \end{eqnarray}
one proves that this kernel is independent of antighosts $\bar C^\pm_\mu$ and satisfies quantum Dirac constraints with respect to both arguments
        \begin{eqnarray}
        \hat H_\mu U(q,q')=0,\quad U(q,q')\overleftarrow{H'}_\mu^{\dagger}=0.
        \end{eqnarray}

This property allows one to interpret $U(q_+,q_-)$ as a kernel of the projector (\ref{projector}) onto the space of quantum Dirac constraints, acting in the representation space of ``matter" variables $\hat q^i,\hat p_i$. In quantum gravity this is a projector onto the space of solutions of Wheeler-DeWitt equations, acting in superspace of 3-metrics and matter fields $q^i=(g_{ab}({\bf x}),\phi({\bf x}))$. This justifies the notation used in Introduction and chosen to serve as the microcanonical density matrix in closed cosmology
        \begin{eqnarray}
        &&U(q,q')=\langle\,q\,|\,\hat U\,|\,q'\,\rangle, \quad
        \hat U=\prod\limits_\mu\delta(\hat H_\mu),\\
        &&\hat\rho=\frac1Z\hat U.
        \end{eqnarray}

Integration over $N_\pm$ in (\ref{U}) implies that this kernel can be interpreted as the unitary evolution kernel in the momentum representation of Lagrange multipliers (\ref{unitarykernel2}) at zero values of $\pi_\pm$,
        \begin{eqnarray}
        &&U(q,q')=\int dN_+ dN_- e^{-i\pi_+N_+}{\mbox{\boldmath$U$}}_{\varPhi}
        (\,t_+,Q_+|\,t_-,Q_-)\;
        e^{i\pi_-N_-}\Big|_{\pi_\pm=C_\pm=0}\nonumber\\
        &&\qquad\qquad=\tilde{\mbox{\boldmath$U$}}_{\varPhi}
        (\,t_+,\tilde Q_+|\,t_-,\tilde Q_-)\,\Big|_{\pi_\pm=C_\pm=0}.   \label{unitarykernel0}
        \end{eqnarray}
Therefore, it has the path integral representation (\ref{unitarykernel1}) with the symbol of the unitarizing Hamiltonian -- the Poisson superbracket of the gauge fermion $\varPhi$ and BRS charge $\varOmega$,
        \begin{eqnarray}
        U(q_+,q_-)
        =\int\limits_{\tilde Q(t_\pm)
        =\tilde Q_\pm} \tilde D[\,Q,P\,]\,\exp\left.
        \Big[\,i\int_{t_-}^{t_+} dt\,
        \big(\tilde P_I\dot{\tilde Q}^I-\{\varPhi,\varOmega\}\big)\Big]\,
        \right|_{\,\pi_\pm=C_\pm=0}.   \label{unitarykernel3}
        \end{eqnarray}
With the conventional choice of the gauge fermion generating the relativistic Faddeev-Popov gauge condition of the form $\dot N^\mu-\chi^\mu(q)=0$ \cite{CERNrep,BFV}
        \begin{eqnarray}
        \varPhi={\cal P}_\mu N^\mu+\bar C_\mu\chi^\mu(q)
        \end{eqnarray}
(note that it differs from the BM gauge fermion (\ref{BMferm})) this path integral becomes
        \begin{eqnarray}
        &&U(q_+,q_-)
        =\int\limits_{\tilde Q(t_\pm)=\tilde Q_\pm} \tilde D[\,Q,P\,]\,%\nonumber\\
        %&&\quad\times
        \exp\Big[\,i\int_{t_-}^{t_+} dt\,\big(\tilde P_I\dot{\tilde Q}^I-N^\mu H_\mu-\pi_\mu\chi^\mu\nonumber\\
        &&\qquad\qquad\qquad\qquad\qquad\qquad\qquad\quad
        -\bar C_\mu J^\mu_\nu\,C^\nu-
        {\cal P}_\alpha(\bar{\cal P}^\alpha+U^\alpha_{\mu\nu}N^\mu C^\nu)
        \big)\Big]\,
        \Big|_{\,\pi_\pm=C_\pm=0}.   \label{1000}
        \end{eqnarray}

A standard procedure of transition to the unitary gauge $\chi^\mu(q)=0$ consists in rescaling the gauge function $\chi^\mu$ by a small numerical parameter $\varepsilon$, $\chi^\mu\to\chi^\mu/\varepsilon$ and making the change of integration variables $\pi_\mu$ and $\bar C_\mu$ (with a unit Jacobian)
        \begin{eqnarray}
        \pi_\mu\to\varepsilon\pi_\mu,\quad \bar C_\mu\to\varepsilon\bar C_\mu.
        \end{eqnarray}
Note that boundary conditions at $t=t_\pm$ admit this change of variables, and all this does not affect the answer for $U(q,q')$ in view of its gauge independence. Then in the limit $\varepsilon\to 0$ the kinetic terms $-N^\mu\dot\pi_\mu$ and $\bar{\cal P}^\alpha \dot{\bar C}_\alpha$ disappear from the integrand of (\ref{1000}), Gaussian integration over ghost momenta does not contribute any field-dependent measure, and the projector to Dirac states takes the form of the usual canonical Faddeev-Popov path integral
        \begin{eqnarray}
        &&U(q_+,q_-)
        =\int D[\,q,p\,]\, DN\,D\pi\,DC\,D\bar C \nonumber\\
        &&\qquad\qquad\quad\times
        \exp\left.\Big[\,i\int_{t_-}^{t_+} dt\,\big(p_i\dot q^i-N^\mu H_\mu-\pi_\mu\chi^\mu -\bar C_\mu J^\mu_\nu\,C^\nu\big)
        \Big]\,\right|_{\,q(t_\pm)=q_\pm,\;\pi_\pm=C_\pm=0}\nonumber\\
        &&\qquad\qquad\quad=\int\limits_{q(t_\pm)=q_\pm} D[\,q,p\,]\,DN\,\Big(\prod\limits_{t_+>t>t_-}\,\delta(\chi)\,\det
        J^\mu_\nu\Big)\exp\Big[\,i\int_{t_-}^{t_+} dt\,\big(p_i\dot q^i-N^\mu H_\mu\big)
        \Big],
        \end{eqnarray}
in which, however, the gauge fixing factor $\delta(\chi)\det
J^\mu_\nu$ is absent at the both end points $t_\pm$. This completely specifies the expression (\ref{rhocanonical}) formulated in Introduction as the microcanonical density matrix and proves all its properties -- projection on the subspace of Dirac constraints (\ref{projU}), gauge independence (\ref{gaugeind}) and independence of the choice of $t_\pm$ (\ref{timeind}). Time here plays merely the role of ordering parameter resolving the issue of noncommutative constraints $\hat H_\mu$.

Statistical sum should be constructed by tracing this kernel with the physical inner product measure
        \begin{eqnarray}
        Z={\rm tr}_{\rm phys}\hat U=
        \int dq\int d\pi\,dC\,d\bar C\,
        e^{-i\bar C_\mu\hat J^\mu_\nu C^\nu+\pi_\mu\hat\chi^\mu}U(q_+,q_-)
        \,\big|_{\,q_-=q_+}.                        \label{trace}
        \end{eqnarray}
This procedure adds the missing gauge fixing factor at the junction of $q_+$ and $q_-=q_+$ , and the trace takes the form of the path integral over periodic histories with $q(t_+)=q(t_-)$
        \begin{eqnarray}
        Z={\rm tr}_{\rm phys}\hat U=\int\limits_{\rm periodic} D[\,q,p\,]\,DN\,\Big(\prod_t\,\delta(\chi)\,\det
        J^\mu_\nu\Big)\exp\Big[\,i\int_{t_-}^{t_+} dt\,\big(p_i\dot q^i-N^\mu H_\mu\big)
        \Big].
        \end{eqnarray}

Now again one can use the gauge independence of this path integral and identically convert it to the BRST integral in relativistic gauges involving Lagrange multipliers and their time derivatives. Integration over momenta then gives by a well known procedure the Lagrangian version of the Faddeev-Popov path integral \cite{CERNrep} over periodic metric and matter field configurations $q^i,N^\mu=(g_{\mu\nu}(x),\phi(x))$. This finally confirms the expression (\ref{Z}).

\section{Conclusions}

The microcanonical density matrix in closed cosmology has a natural definition as a projector on the space of solutions of Wheeler-DeWitt equations, which is motivated by the absence of global non-vanishing charges in spatially closed gravitational systems. However, the very definition of this projector encounters difficulties because the quantum Dirac constraints -- the operators of Wheeler-DeWitt equations -- do not commute and form an open algebra in contrast to the case of closed compact algebra, when it can be achieved by group integration. For this reason we have built this projector indirectly via the construction of the unitary evolution operator in the BRST/BFV formalism of generic gauge systems. Its truncation to the sector of ``matter" variables yields this projector in the form of the canonical Faddeev-Popov path integral over a special class of histories interpolating between the arguments of its kernel in the (functional) coordinate representation. The ``time" variable parameterizing this histories plays the role of auxiliary operator ordering parameter, resolving the issue of noncommutative quantum constraints. This property is in full accordance with the fact that in time parametrization invariant invariant systems (like spatially closed cosmology) the BFV unitary evolution operator represents evolving BRS gauge transformation.

Truncation of the BRST wavefunction in the representation space of ``matter" and ghost variables to the matter sector is in fact the Batalin-Marnelius gauge fixing which selects physical BRS  siglets in this space. This leads to the regularization of the physical inner product in the form of a special integration measure in the inner product of the Dirac quantization. In fact this is a quantum measure which provides the composition law for the Faddeev-Popov path integral. When applied to the density matrix, the physical inner product gives its statistical sum as the Lagrangian path integral over periodic configurations of spacetime metric and matter fields (\ref{Z}).

BRST aspects of the density matrix and statistical sum, of course, extend beyond the closed cosmology setup. They equally apply to black hole systems with asymptotically flat and horizon boundaries which are responsible for non-vanishing global charges including the energy. The associated boundary terms serve as a non-vanishing Hamiltonian $H_0(q,p)$ of (\ref{1.1}). It drives the physical evolution which is not reducible to pure gauge transformations and provides a conventional definition of the microcanonical statistical sum as a function of energy $E$. The latter has a path integral representation (\ref{Z}) with the action including these surface terms subject to boundary conditions matching with the value of $E$ \cite{BrownYork}. In particular, the time period of its configurations is determined by this value -- a free argument of the microcanonical state with the energy $E$ (the Legendre transform with respect to $E$ gives the density matrix of the canonical ensemble with the temperature given by the imaginary value of this period). This is different from spatially closed cosmology (\ref{rhocanonical}) whose microcanonical density matrix is independent of $t_\pm$ and obviously is not labeled by the energy value.

Important problem of the BRST formalism is admissibility of gauge fixing procedure which should be globally applicable in configuration space. Gauge conditions with globally non-degenerate Faddeev-Popov operator are not known (being, perhaps, not available at all) and give rise to the Gribov copies problem. In context of the diffeomorphism invariance this has a manifestation of the so-called ``problem of time" \cite{BarvU}. This problem is usually disregarded within perturbation theory, though even in semiclassical expansion it arises at the caustics of congruences of classical histories \cite{UFN}, where the junction of the Lorentzian domain with the classically forbidden (Euclidean quantum gravity) domain takes place. In particular, it poses the dilemma of the no-boundary \cite{no-boundary} vs tunneling \cite{tunneling} cosmological states. The formulation of the initial conditions for the early Universe in the form of the microcanonical density matrix \cite{why,slih} suggests a serious alternative to this dilemma, but does not resolve the gauge fixing issue which will be considered elsewhere.

Of course, the BRST aspects of quantum cosmology are not exhausted by the above considerations. The choice of definite vs indefinite metric quantization in the sector of matter variables $(q^i,N^\mu)$, discussed in \cite{rules}, remains important even semiclassically because it determines convexity properties of the gravitational action at the saddle points of the path integral (\ref{Z}) and serves for the selection of cosmological instantons \cite{slih}. Also it is noteworthy that specifics of the cosmological setting (except its time parametrization invariance) was never used in the above formal derivations which apply to generic gauge systems. At the same time, the assumption of Hermiticity of matter operators $(\hat q^i,\hat p_i)$ on the space of Dirac wavefunctions is a corner stone of the BRST/BFV formalism, and it requires verification at least in simple minisuperspace models of quantum cosmology \cite{selection}.

\section*{Acknowledgements}
I am grateful to I.V.Tyutin for helpful discussions and
wish to express my gratitude for hospitality of the Yukawa Institute for Theoretical Physics during the workshop YITP-T-12-03 ``Gravity and Cosmology 2012" where this work was initiated. This work was also supported by the RFBR grant No 11-01-00830.

\end{document}